\documentclass[11pt]{article}
\usepackage{moriond}
\usepackage{epsfig}
\usepackage{wrapfig}
\usepackage{amsmath}
\usepackage{tabularx}

\def\Journal#1#2#3#4{{#1} {\bf #2}, #3 (#4)}

\def\APJ{\em ApJ}

\newcommand{\ea}{et~al.\ }
\newcommand{\pr}{\textrm{Pr}}
\newcommand{\data}{\textrm{\,Data\,}}
\newcommand{\model}{\textrm{\,Model\,}}
\newcommand{\pars}{\,\{\theta\}\,}
\newcommand{\Ob}{$\Omega_B$}
\newcommand{\Om}{$\Omega_m$}
\newcommand{\ob}{\Omega_B}
\newcommand{\om}{\Omega_m}

\begin{document}
\vspace*{4cm}
\title{COSMOLOGY FROM CLUSTER SZ AND WEAK LENSING DATA}

\author{ P.J. Marshall, K. Lancaster, M.P. Hobson and K.J.B. Grainge}

\address{Astrophysics Group, Cavendish Laboratory, Madingley Road,
Cambridge CB3 0HE, England}

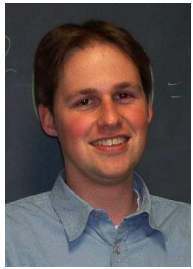
\maketitle\abstracts{ Weak gravitational lensing and the
Sunyaev-Zel'dovich effect provide complementary information on the
composition of clusters of galaxies. 
Preliminary results from cluster SZ observations with the Very Small Array
are presented.  A Bayesian approach to combining this data with wide
field lensing data is then outlined; this allows the relative
probabilities of cluster models of varying complexity to be computed. A
simple simulation is used to demonstrate the importance of cluster
model selection in cosmological parameter determination.}

\section{Introduction}
\enlargethispage{2\baselineskip}

As the most massive gravitationally bound
structures in the Universe, clusters of galaxies
are powerful probes of cosmology.  However, to be able to extract
reliable cosmological information from cluster data,
their astrophysical structure must first be understood.  For example,
one assumption often made about clusters is that of the intra-cluster
gas being in hydrostatic equilibrium with the gravitational potential
defined by the cluster dark matter distribution. Any cosmological
conclusions drawn are then dependent on the validity of this
assumption.

Historically clusters have been studied primarily in the X--ray and
optical wavebands, but in recent years the observation of cluster
atmospheres via the Sunyaev-Zel'dovich effect,\cite{Bir99} and cluster
potentials through weak gravitational lensing,\cite{KS93,Mar++01} have
become routine. These two methods are complementary, in
the following way. Measurement of the statistical distortions of
background galaxy ellipticities due to gravitational lensing allows the
projected  distribution of total mass to be inferred, while mapping the
decrement in the microwave surface brightness of the CMB
provides a measure of the projected gas pressure. 
From these two observations,
and some knowledge of the cluster's temperature, one might hope to be
able to compare the cluster gas mass and total mass, and also to assess
the degree to which the cluster is in equilibrium.

Presented here are some preliminary results of observations of
low redshift
clusters with the Very Small Array~(VSA). We then
outline our approach to combining such data with that expected from
wide-field lensing observations, and give a simple
demonstration of the cosmological results that may be obtained.
\newline
\begin{wrapfigure}[18]{r}{8.5cm}
\centering\epsfig{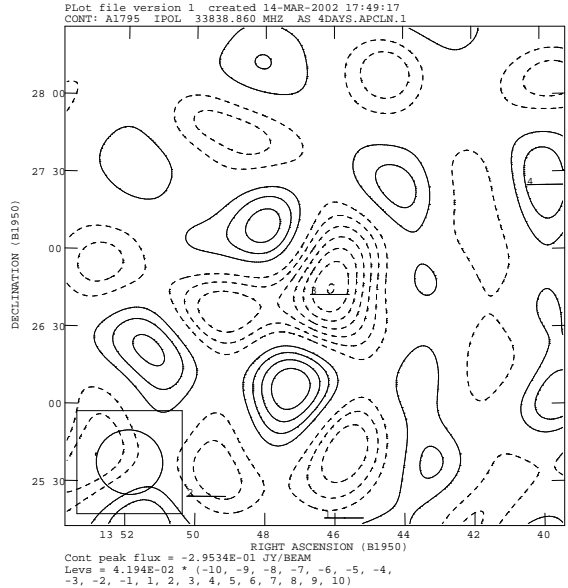} \caption{SZ
map of Abell 1795. The secondary features around the cluster are
primordial CMB fluctuations.} \label{fig:vsa} \end{wrapfigure}

\vspace{-3\baselineskip}
\section{Results from the VSA}
\enlargethispage{\baselineskip}

The VSA is a fourteen element interferometer operating at 34~Ghz
situated at the
Observatorio del Teide on Tenerife~\cite{VSA}. In
its current extended antenna configuration the VSA is able to image the
microwave sky on angular scales between and 9 arcminutes and 2 degrees 
($1200>l>400$); a rich cluster at a redshift of 0.07 has a virial radius 
of around 0.5 degrees. The VSA is in the process of observing a sample
of six clusters selected for their X-ray brightness: a map of A1795 from
20 hours observation is given in Figure~\ref{fig:vsa}. As the
figure shows, the cluster is resolved, suggesting that more than just a
single parameter describing the temperature decrement may be determined
from this data.

\vspace{2\baselineskip}
\section{Bayesian Parameter Fitting and Model Selection}

Fitting simple physically-motivated parameterised models to cluster data
allows model--dependent statements about cosmology to be made; 
here we extend this current standard
to include a second level of inference, asking ``Which of our proposed
models is most probable, given all available data?''  To answer this 
question we should calculate
\begin{equation*}
  \pr\,(\model|\data) \propto \pr\,(\data|\model)\; \pr\,(\model).
\end{equation*}
\noindent If we have no prior preference for one model over another, we
can express the answer to the above question as the ratio
$\pr\,(\data|\model_1)\,/\,\pr\,(\data|\model_2)$.
The value of the ``evidence'' $\pr\,(\data|\model)$ may be
calculated by marginalising the likelihood  $\pr\,(\data|\pars,\model)$
over the $N$~free parameters~$\pars$ of the model in
question. The evidence is the denominator of Bayes' Theorem in the
context of investigating the posterior probability of the parameters of
the model,
\begin{equation*}
  \pr\,(\pars|\data,\model) = \frac{\pr\,(\data|\pars,\model)\; \pr\,(\pars|\model)}
	                                {\pr\,(\data|\model)} .
\end{equation*}

Therefore, the procedure is to model the data and their
errors to construct the likelihood,
multiply by the appropriate  prior probability distribution for each
parameter (\emph{e.g.} $1<T/\rm{keV}<10$ with uniform probability) and
then calculate the posterior $\pr\,(\pars|\data,\model)$, which can
be normalised to find the evidence. Estimators for the parameters
of the model may be taken as those which maximise the posterior
distribution, which also provides confidence intervals for the
assessment of parameter  uncertainties.

In practice even the simplest cluster models will have around
8 free parameters -- calculating the posterior probability distribution
on even a very coarse 10 pixel grid would take of order several years 
of computing time. Conversely, finding the ``best--fit'' points by
numerically maximising the posterior probability is fast, but this
procedure is
prone to getting caught in one of the many local maxima present in the space,
and does not provide information on the parameter uncertainties.

An efficient way to investigate the posterior is by using an
exploratory Markov Chain Monte-Carlo algorithm.\cite{mcmc}  This
process builds up an ensemble of sample points in the $N$-dimensional
parameter space, preferentially sampling in regions where the posterior
density is high. After convergence of the algorithm the final ensemble
of samples (whose number density is proportional to the posterior
density) may then be used to construct any required statistics.

\section{Joint Analysis of Simulated Low Redshift Cluster Observations}

To demonstrate our cluster analysis under controlled conditions
realistic datasets were generated, and the model parameters
investigated. 

Firstly, the SZ data are the visibilities $Re(V_i)$ and $Im(V_i)$
where the $V_i$ are noisy samples of the Fourier transform of the
microwave sky surface brightness $I_{\nu}$. Typical VSA
observing parameters were used, and
Gaussian noise appropriate to a 30 hour observation added. The
likelihood for this data is 
\begin{equation*}
 \log{\pr (\data|\pars,\model)} =  
	\sum  \left[ \log{\frac{1}{\sqrt{2 \pi \sigma_{V}^2}}} - 
       \frac{(V_i - FT(I_{\nu}))^2}{2 \sigma_{V}^2} \right]
\end{equation*}
Inferring the five model parameters of a spherically symmetric 
isothermal beta model \newline 
\mbox{$\{x,y,T \cdot \rho_{gas,0}, r_c, \beta\}$}, 
indicated that  although the core radius and the~$\beta$
parameter may be only poorly constrained by the VSA data, the total gas
mass may be better estimated, as found elsewhere.\cite{Gre++01} 

A similar test was performed for weak gravitational lensing data. The 
field of view of 
MegaCam at the CFHT is very well matched to the VSA cluster 
observations,
and provides enough background galaxies to make up for the non-optimal
lensing geometry at the VSA cluster redshifts.
A dataset corrsponding to a 2 hour observation 
was simulated for a spherically symmetric NFW
profile cluster (4 parameters, $\{x,y,r_s,\rho_s\}$).
The likelihood in this
case is
\begin{equation*}
 \log{\pr (\data|\pars,\model)} = \sum \left[ \log{\frac{1}{\sqrt{2 \pi \sigma_{\epsilon}^2}}} - 
       \frac{|\epsilon - g|^2}{2 \sigma_{\epsilon}^2} \right]
\end{equation*}
\noindent where $\epsilon_1$ and $\epsilon_2$ are the background galaxy
ellipticity parameters, assumed to sample (noisily) the reduced shear
field $g$ which can be calculated for a given mass distribution. 
As in the SZ data case
the constraints on the
individual parameters were not found to be very tight -- however, 
the data
did allow the total mass of the cluster to be measured to 
within 25 percent.  
To investigate the relative probabilities of a set of models describing
both the gas and total mass distributions,
given both
weak lensing and Sunyaev-Zel'dovich data, the joint
likelihood can be constructed by just multiplying the likelihoods 
together.
Integrating the model gas and total mass density profiles to a given
radius 
allows the cluster gas fraction to be calculated;
if we make the ``fair sample'' assumption and
normalise the gas mass such that  $\frac{M_{gas}}{M_{total}} =
\frac{\ob}{\om}h$,  we can include \Om and \Ob\; as extra parameters in
our model and \emph{learn about the cluster and cosmology at the same
time.} Combining the results from the Hubble Key Project\cite{HSTkey}
and from Big Bang Nucleosynthesis\cite{BBN} gives the Gaussian prior
$\ob = (0.04 \pm 0.01)$, while we assume a uniform prior on~\Om\; over 
the range 0.1 to 1.  

As a toy demonstration the data sets described above (representing a
cluster which is \emph{not} in hydrostatic equilibrium) were fitted
with two models, the true distributions (Model 1, 9
parameters, $\{x, y, r_s, \rho_s, T_0, r_c, \beta, \om, \ob\}$), and an
equilibrium gas distribution in an NFW potential, with allowance for a
temperature drop in the core (Model 2, 9 parameters,
$\{x, y, r_s, \rho_s, T_0, T_1,$ $ \frac{d \log T}{d \log
r}|_{\rm{core}}, \om, \ob\}$).

\begin{figure}[!t]
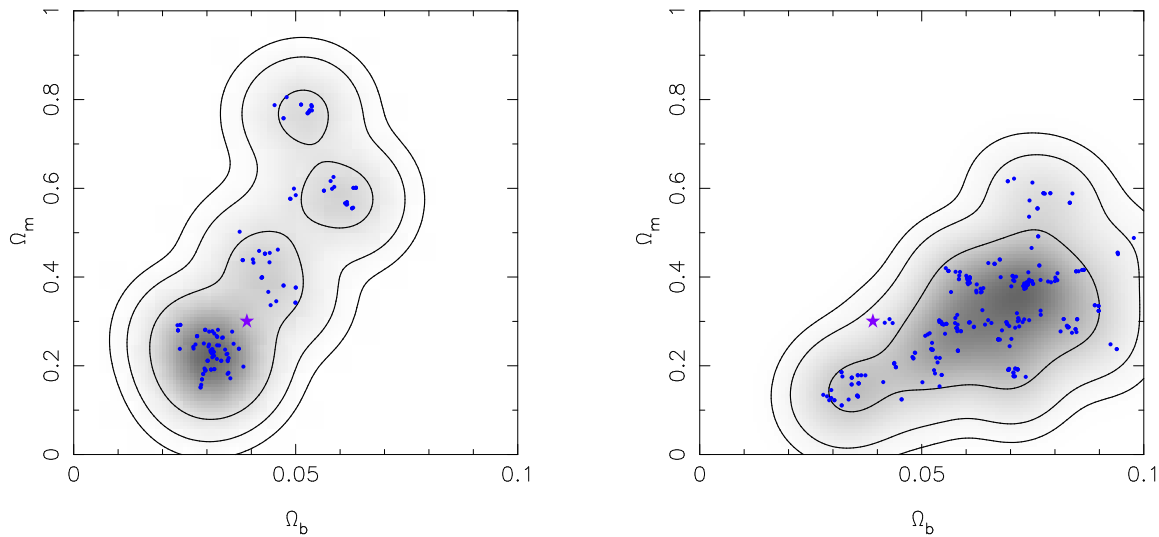

\begin{minipage}[b]{0.48\linewidth}
\centering\epsfig{file=cosmo_beta.ps,width=0.9\linewidth,angle=0,clip=}
\end{minipage} \hfill
\begin{minipage}[b]{0.48\linewidth}
\centering\epsfig{file=cosmo_equi.ps,width=0.9\linewidth,angle=0,clip=}
\end{minipage}
\caption{Cosmological parameters inferred for the two cluster models
outlined in the text (Left, Model1; Right, Model2). The ``true'' values
are
marked with a star, contours represent 68, 95 and 99 percent confidence
intervals. The samples used in generating the smooth distributions are
shown as points.}
\label{fig:cosmo}
\end{figure}

The cosmological results from these two analyses are shown in
Figure~\ref{fig:cosmo}, where all other 
\clearpage
\noindent parameters have been
marginalised over. The model-dependence of the conclusions about the 
most probable values of~\Om \,and~\Ob \,can be clearly seen. Despite
having very similar reduced chi-squared values, the
evidence ratio for the two models is in favour of the true cluster
model:
\begin{equation*} 
\frac{ \pr\,(\data|\model_1) }{ \pr\,(\data|\model_2) } = 11. 
\end{equation*}

\section{Conclusions and Future Work}
\enlargethispage{3\baselineskip}

The Bayesian approach to cluster data analysis 
presented here is the natural
extension of methods currently employed by observers: it provides an
efficient way of exploring high-dimensional parameter spaces, a direct
route from the experimental noise to uncertainties on the model
parameters, and, in calculating the evidence, a way of discriminating
between model-dependent statements.

We intend to include a larger array of cluster
models of varying complexity, appropriate to the ever-improving data
such as that expected from  the next generation of SZ telescopes. 
X-ray data can also be included straightforwardly into the analysis,
providing information on~$H_0$ and the cluster temperature as well as
improving the constraints on the other parameters. 

\section*{Acknowledgments}
PJM and KL acknowledge support in the form of PPARC studentships.

\end{document}